%% file: template.tex
\RequirePackage{fix-cm}
\documentclass[smallcondensed, numbook]{svjour3}     
\smartqed  

\usepackage[table]{xcolor}
\usepackage{tikz}
\usepackage{chngpage}
\usetikzlibrary{shapes.geometric, arrows}

\newcommand\Tau{\mathcal{T}} 
\usepackage[acronym]{glossaries}
\usepackage{booktabs}
\usepackage{threeparttable}
\usepackage{siunitx}
\usepackage{multirow}

\usepackage[round]{natbib} 

\usepackage{graphicx}
\usepackage{amsmath}
\usepackage{amssymb}
\usepackage[naturalnames]{hyperref}
\usepackage{cleveref}

\input{acronyms}
\graphicspath{ {figures/} }

\tikzstyle{node} = [rectangle, rounded corners, minimum width=3cm, minimum height=0.5cm, text centered, draw=black]
\tikzstyle{arrow} = [thick,->,>=stealth]
\tikzset{font={\fontsize{8pt}{12}\selectfont}}

\journalname{Currently Under Review}

\begin{document}

\title{A hydraulic model outperforms work-balance models for predicting recovery kinetics from intermittent exercise}

\author{
Fabian C. Weigend\hspace{1mm}\href{https://orcid.org/0000-0001-8868-9735}{\includegraphics[scale=0.06]{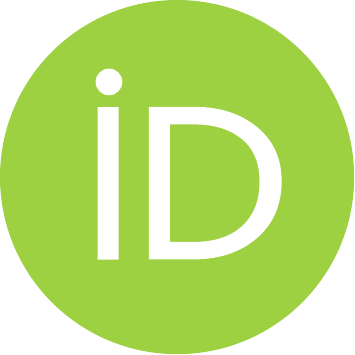}}
\and
David C. Clarke\hspace{1mm}\href{https://orcid.org/0000-0002-1520-5426}{\includegraphics[scale=0.06]{orcid.pdf}}
\and
Oliver Obst\hspace{1mm}\href{https://orcid.org/0000-0002-8284-2062}{\includegraphics[scale=0.06]{orcid.pdf}}  
\and
Jason Siegler\hspace{1mm}\href{https://orcid.org/0000-0003-1346-4982}{\includegraphics[scale=0.06]{orcid.pdf}}
}

\authorrunning{Weigend et al.} 
\titlerunning{A hydraulic model for predicting recovery kinetics}

\institute{
        Fabian C. Weigend \at
            School of Computer, Data and Mathematical Sciences and School of Health Sciences \\
            Western Sydney University\\
            \email{Fabian.Weigend@westernsydney.edu.au}
        \and
        David C. Clarke \at
            Department of Biomedical Physiology and Kinesiology and the Sports Analytics Group \\
            Simon Fraser University
        \and
        Oliver Obst \at
            School of Computer, Data and Mathematical Sciences\\
            Western Sydney University
        \and
        Jason Siegler \at
            College of Health Solutions\\
            Arizona State University
}

\date{Received: date / Accepted: date}

\maketitle

\begin{abstract}

Data Science advances in sports commonly involve ``big data'', i.e., large sport-related data sets. However, such big data sets are not always available, necessitating specialized models that apply to relatively few observations. One important area of sport-science research that features small data sets is the study of recovery from exercise. In this area, models are typically fitted to data collected from exhaustive exercise test protocols, which athletes can perform only a few times. Recent findings highlight that established recovery such as the so-called work-balance models are too simple to adequately fit observed trends in the data. Therefore, we investigated a hydraulic model that requires the same few data points as work-balance models to be applied, but promises to predict recovery dynamics more accurately.

To compare the hydraulic model to established work-balance models, we retrospectively applied them to data compiled from published studies. In total, one hydraulic model and three work-balance models were compared on data extracted from five studies. The hydraulic model outperformed established work-balance models on all defined metrics, even those that penalize models featuring higher numbers of parameters. These results incentivize further investigation of the hydraulic model as a new alternative to established performance models of energy recovery.

\end{abstract}

\section{Introduction}
\label{intro}

Emerging technologies that enable the real-time monitoring of athletes in training and competition have fostered interest in methods to predict and optimize athlete performance. Predictive models for how much an athlete ``has left in the tank'' enable the investigation of pacing strategies \citep{behncke_optimization_1997,sundstrom_comparing_2014,de_jong_individual_2017} and to dynamically adjust strategies to optimize the outcome of a competition \citep{hoogkamer_modeling_2018}. They can be described as a digital athlete, i.e., a computer-based model for enhancing training programming or strategy optimization. A foundation for such advances is research in performance modeling, which can be understood as the mathematical abstraction of exercise physiology.

\subsection{Critical-power-based approaches to performance modeling}
\label{subsec:current_app}

One of the seminal models in the area of performance modeling is the critical power model, which relies on the notions of a \gls*{cp} and a \gls*{w'} \citep{hill_critical_1993, whipp_constant_1982}. \citet{monod_work_1965} defined \gls*{cp} as: ``the maximum work rate a muscle can keep up for a very long time without fatigue''. Thus, \gls*{cp} can be considered as a threshold for sustainable exercise. \gls*{w'} represents a capacity for work to be performed at a rate above \gls*{cp} and is conceptualized as an energy storage. Using the definitions of \gls*{tte} and \gls*{p} the critical power model can be summarized in the relationship

\begin{equation}\label{eq:cp_model}
    \gls*{tte} = \frac{\gls*{w'}}{\gls*{p}-\gls*{cp}}.
\end{equation} 

To determine \gls*{cp} and \gls*{w'}, an athlete has to conduct between three and five exercise tests until exhaustion at various constant exercise intensities. \gls*{cp} and \gls*{w'} are then fitted to these distinct \gls*{tte} and \gls*{p} observations and their relationship can be used to predict the time to exhaustion at other intensities. \citet{hill_critical_1993} emphasized that the attractiveness of the model lies in its coarse simplicity and it should not be employed if highly accurate predictions are required. Nevertheless, its straightforward application and its elegant abstraction have led to improved understanding and prediction of performance dynamics \citep{poole_critical_2016,sreedhara_survey_2019,vanhatalo_application_2011}.

While the critical power model predicts energy expenditure at high intensities, it does not consider the recovery of \gls*{w'} after exercise has ended or during exercise at low intensities. Formally, exercise protocols that alternate between intensities below \gls*{cp} and above \gls*{cp} constitutes as intermittent exercise. In order to predict performance capabilities of athletes during intermittent exercise, models need to predict recovery of \gls*{w'} during phases of exercise below the \gls*{cp} intensity. 

One of the most widely covered approaches to predict recovery of \gls*{w'} during intermittent exercise is the \gls*{w'bal} model \citep{sreedhara_survey_2019,jones_critical_2017}. Since the first publication by \citet{skiba_modeling_2012}, an updated form of \gls*{w'bal} was introduced by \citet{skiba_intramuscular_2015} and another alternative form was proposed by \citet{bartram_accuracy_2018}. \gls*{w'bal} models have been used to search for optimal drafting strategies in running \citep{hoogkamer_modeling_2018} or to predict phases of perceived exhaustion during cycling exercise \citep{skiba_validation_2014}. 

Despite these advances, research into energy recovery modeling is an evolving field, and \gls*{w'bal} models have been scrutinized for their limitations. Similar to energy expenditure dynamics, recovery dynamics are derived from exhaustive exercise tests and therefore data for model fitting and validation are sparse \citep{vanhatalo_application_2011,sreedhara_survey_2019}. Recent findings suggest that current \gls*{w'bal} models overly simplify energy recovery dynamics, and that model modifications that account for characteristics of prior exhaustive exercise \citep{caen_reconstitution_2019} as well as bi-exponential recovery dynamics~\citep{caen_w_2021} might improve recovery predictions. These modified models feature additional parameters, which introduces challenges in fitting them to small data sets. Indeed, the search for models that optimally balance complexity with applicability to few data points is a primary challenge of energy recovery modeling.

\subsection{Hydraulic models of human performance}
\label{subsec:hyd}

Hydraulic models offer an alternative to \gls*{w'bal} models for predicting energy expenditure and recovery dynamics during exercise. Instead of using \gls*{cp} and \gls*{w'}, they represent energy dynamics as liquid flow within a system of tanks and pipes. These tanks and pipes are arranged according to physiological parameters such as maximal oxygen uptake and estimated phosphocreatine levels of an athlete. The first hydraulic model was proposed by \citet{margaria_biomechanics_1976} to provide an intuitive conceptualization of bioenergetic responses to exercise. \citet{morton_three_1986} further elaborated the model, formalized its dynamics with differential equations, and published it as the Margaria-Morton (M-M) model. Later, \citet{sundstrom_bioenergetic_2016} proposed an extension of the M-M model and named it the Margaria-Morton-Sundstr\"om (M-M-S) model. Compared to \gls*{w'bal} models, hydraulic models can predict more complex energy expenditure and recovery dynamics and have the potential to address highlighted shortcomings of \gls*{w'bal} models in recent literature. 

A challenge of these hydraulic models is that their parameters require in-depth knowledge about bioenergetic systems. Indeed, \citet{morton_critical_2006} concluded that it remained to be seen to what extent model predictions conform to reality. Also, the more recently proposed \mbox{M-M-S} model by \citet{sundstrom_bioenergetic_2016} has yet to be validated experimentally. \citet{behncke_optimization_1997} applied the \mbox{M-M} model to world records in competitive running, and while the predictions agreed with values provided in the literature, he also pointed out situations in which the naive interpretation of the model would not be justified. Furthermore, \citet{behncke_optimization_1997} stated that constraints dictated by physiological conditions made explicit computations with the \mbox{M-M} model ``rather cumbersome''. Collectively, the requirement to set parameters according to physiological measures impede the application and validation of the M-M and M-M-S models.

To overcome the issues caused by ascribing the model parameters to concrete bioenergetic measures, we proposed a generalized form of the M-M hydraulic model in \citet{weigend_new_2021}. Our generalized hydraulic model allows the fitting of its parameters using an optimization approach that only requires \gls*{cp} and \gls*{w'} as inputs. In this way, our modified model preserves the flexibility needed to model the observed dynamics without requiring strict correspondence to parameters pertaining bioenergetics. In a proof-of-concept, we showed that our fitted hydraulic model could successfully predict both energy expenditure and recovery kinetics for one example case, the former in line with predictions of the \gls*{cp} model and the latter in a manner that matches published observations. While the generalized hydraulic model demonstrated satisfactory predictivity, it is still unknown as to whether it can outperform the existing \gls*{w'bal} models.

Therefore, in this work, we compare the prediction quality of our generalized hydraulic model from \citet{weigend_new_2021} to that of three \gls*{w'bal} models. We hypothesized the hydraulic model would predict the observed recovery ratios compiled from past studies overall more accurately than the \gls*{w'bal} models. We found that the hydraulic model outperformed the \gls*{w'bal} models on objective goodness-of-fit and prediction metrics. We conclude that the generalized hydraulic model provides a beneficial new perspective on energy recovery modeling that should be investigated further.

\section{Material and methods}
\label{sec:material_and_methods}

The \emph{Materials and methods} are structured in the following way. In \Cref{subsec:model_definitions}, the \gls*{w'bal} and hydraulic models are defined and their underlying assumptions specified. We then introduce a new \gls*{w'bal} model that has been fitted to the same recovery data as the investigated hydraulic model. In \Cref{subsec:compare_procedure}, we discuss how we will objectively compare the \gls*{w'bal} and hydraulic models. In particular, we propose a procedure to obtain comparable recovery ratios. The validation data set consists of data compiled from previously published studies on the recovery from exercise. \Cref{subsec:comparison_studies} lists the data exclusion criteria and extraction procedures. Finally, in \Cref{subsec:metrics_of_fit}, we describe the metrics used to assess the model goodness-of-fits and prediction capabilities.

\subsection{Model definitions}
\label{subsec:model_definitions}

The \gls*{cp}, \gls*{w'bal}, and the hydraulic models feature assumptions and parameters that require defining.

\subsubsection{Energy expenditure and recovery with \gls*{w'bal}}
\label{subsec:wbal}

The critical power model predicts energy expenditure and is the underlying model for the \gls*{w'bal} models. The four essential assumptions of the critical power model are stated as follows \citep{hill_critical_1993,morton_critical_2006}:
\begin{itemize}
    \item[1.] An individual's power output is a function of two energy sources: aerobic (using oxidative metabolism) and anaerobic (non-oxidative metabolism).
    \item[2.] Aerobic energy is unlimited in capacity but its conversion rate into power output is limited (\gls*{cp}).
    \item[3.] Anaerobic energy is limited in capacity (\gls*{w'}) but its conversion rate is unlimited.
    \item[4.] Exhaustion occurs when all of \gls*{w'} is depleted.
\end{itemize}
These assumptions are reflected in \Cref{eq:cp_model}, in which time to exhaustion (TTE) is estimated from the available \gls*{w'} divided by work above \gls*{cp}. At every time step during which an athlete exercises above \gls*{cp}, the product of the time elapsed and the difference between \gls*{cp} and the actual power output is subtracted from the energy capacity \gls*{w'}. Thus, during a constant power output above \gls*{cp}, the critical power model predicts a linear depletion of \gls*{w'}. When \gls*{w'} is depleted, exhaustion is reached.

\begin{figure}[b]
    \centering
    \includegraphics[width=0.8\linewidth]{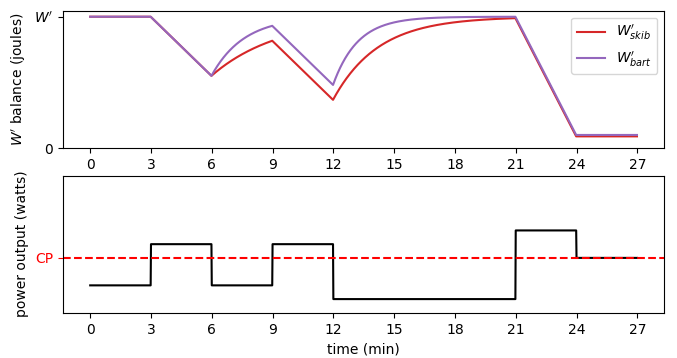}
    \caption{Example energy expenditure and recovery predictions of two models for intermittent exercise. \gls*{w'} expenditure during exercise above \gls*{cp} (the red dashed line in the lower panel) was modeled using the \gls*{cp} model while \gls*{w'} recovery during exercise below \gls*{cp} was modeled using either the \gls*{w'bal-skib} or \gls*{w'bal-bart}. Differences in predicted recovery kinetics are clearly visible.}
    \label{fig:w_bal_sim}
\end{figure}

As observable in the example in \Cref{fig:w_bal_sim}, subsequently established \gls*{w'bal} models combine the assumed linear depletion of \gls*{w'} at power outputs above \gls*{cp} with predictions for \gls*{w'} reconstitution during exercise below \gls*{cp}. The initial \gls*{w'bal} model by~\citet{skiba_modeling_2012} was later updated in~\cite{skiba_intramuscular_2015}. Substantial differences between these versions exist, and as shown by \citet{skiba_w_2021}, the original model by~\citet{skiba_modeling_2012} contradicts the assumption of \Cref{eq:cp_model} that \gls*{w'} linearly depletes. As such, we focused on the updated version by~\citet{skiba_intramuscular_2015} and we refer to it henceforth as \gls*{w'bal-ode}. We denote the remaining capacity of \gls*{w'} at a discrete time point $t$ during exercise as $W^\prime_{\mathrm{bal-ode}_{t}}$. $P_t$ refers to the power output at a discrete time step $t$. $\Delta t$ is the difference between the discrete time step $t-1$ and $t$ in seconds.  We define the overall \gls*{w'bal-ode} model as
\begin{equation}\label{eq:differential_rec}
  W^\prime_{\mathrm{bal-ode}_{t}} =
  \begin{cases}
    W^\prime_{\mathrm{bal-ode}_{t-1}} - (P_t - \gls*{cp})\Delta t, & \text{for} \; P_t \geq \gls*{cp}\\
    \gls*{w'} - (\gls*{w'} - W^\prime_{\mathrm{bal-ode}_{t-1}}) \cdot e^{\frac{-\Delta t}{\Tau_t}}, & \text{for} \; P_t < \gls*{cp}. \\
  \end{cases}
\end{equation}
During a constant power output above or at \gls*{cp} ($P_t \geq \gls*{cp}$), $W^\prime_{\mathrm{bal-ode}_{t}}$ decreases linearly as $t$ increases, like the critical power model predicts. During power outputs below \gls*{cp}, $W^\prime_{\mathrm{bal-ode}_{t}}$ increases exponentially with \gls*{w'} as its asymptote. $\Tau_t$ affects recovery speed and varies between distinct \gls*{w'bal-ode} models. At a discrete time step $t$, the \gls*{tau-skib} is estimated as
\begin{equation}\label{eq:skiba_2015_tau}
     \Tau_{\mathrm{skib}_t} = \frac{\gls*{w'}}{D_{\gls*{cp}_t}},
\end{equation}
where $D_{\gls*{cp}_t}$ represents the difference between $P_t$ and \gls*{cp}. Henceforth, we refer to \Cref{eq:differential_rec} with \Cref{eq:skiba_2015_tau} as \gls*{w'bal-skib}. \Cref{fig:w_bal_sim} depicts an example for \gls*{w'bal-skib} predictions. For the time steps of the first 3 minutes $P_t$ was below \gls*{cp} and $W^\prime_{\mathrm{bal-ode}_{t}}$, i.e., available $W^\prime$ balance, remained at its maximum. Then, the power output increased above \gls*{cp} for the next three minutes. The available \gls*{w'} balance decreased by $P_t - \gls*{cp}$ per second. Between 6 and 9 minutes $P_t$ dropped below \gls*{cp} again, \gls*{w'bal-skib} simulated recovery, and $W^\prime_{\mathrm{bal-ode}_{t}}$ rose exponentially with \gls*{w'} as its asymptote. Speed of recovery was affected by $\Tau_{\mathrm{skib}_t}$ from \Cref{eq:skiba_2015_tau}, which took the difference between $P_t$ and \gls*{cp} into account. That is observable by comparing recovery between 6 and 9 minutes to recovery between 12 and 21 minutes. During the second recovery bout $P_t$ was lower and therefore the slope of the exponential recovery was steeper. During the last three minutes $P_t$ was equal to \gls*{cp} and thus $W^\prime_{\mathrm{bal-ode}_{t}}$ did not change. If $W^\prime_{\mathrm{bal-ode}_{t}}$ would reach 0, exhaustion would be predicted. In the example in \Cref{fig:w_bal_sim} the athlete was predicted to be close to exhaustion, but some of their energy capacities remained.

\citet{bartram_accuracy_2018} investigated the recovery rate of \gls*{w'} of elite cyclists and observed faster recovery rates than \citet{skiba_intramuscular_2015}. Therefore, they proposed another $\Tau_t$ for \Cref{eq:differential_rec} to predict quicker recovery ratios. The \gls*{tau-bart} was defined as
\begin{equation}\label{eq:bartram_tau}
    \Tau_{\mathrm{bart}_t} = 2287.2 \cdot {D_{\gls*{cp}_t}}^{-0.688}.
\end{equation}
Henceforth, we will refer to \Cref{eq:differential_rec} with \Cref{eq:bartram_tau} as \gls*{w'bal-bart}. Predictions of \gls*{w'bal-bart} are depicted alongside those of \gls*{w'bal-skib} in the example in \Cref{fig:w_bal_sim}. It is observable that \gls*{w'bal-bart} predicted faster recovery dynamics.

\subsubsection{The hydraulic model}
\label{subsec:hyd_model}

\begin{figure}
    \begin{center}
    \includegraphics[width=0.8\linewidth]{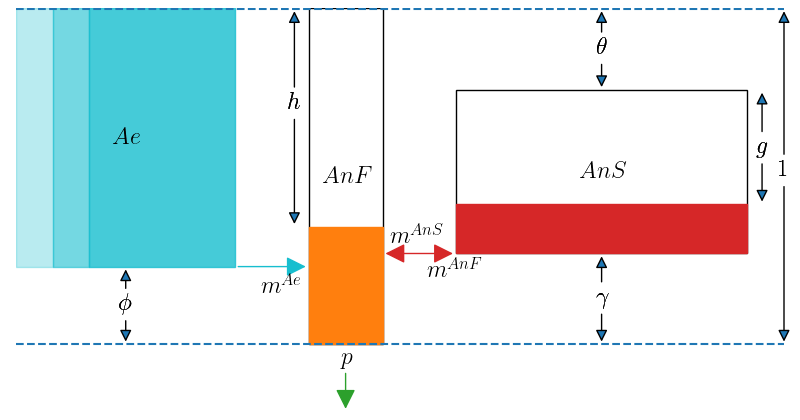}
    \caption{A three-component hydraulic model as defined by~\cite{weigend_new_2021}. Tanks are conceptualized as the aerobic energy source ($Ae$), anaerobic fast energy source ($AnF$), and anaerobic slow energy source ($AnS$). $Ae$ is assumed to be infinite in volume, which is indicated by the fading color to the left. A tap $p$ is attached to the bottom of $AnF$ and flow from it represents energy demand. Pipes connecting the three tanks have maximal flow capacities $m^{Ae}$, $m^{AnS}$, and  $m^{AnF}$.}
    \label{fig:three_comp}
    \end{center}
\end{figure}

In \citet{weigend_new_2021} we expressed our generalized hydraulic tank model mathematically as a system of discretized differential equations with 8 parameters (A.3 and A.4 in their Appendix). Henceforth, the model will be referred to as the \gls*{hyd-weig}, a schematic of which is depicted in \Cref{fig:three_comp}. \Gls*{hyd-weig} models power output of an athlete as a function of three interacting energy sources, which are represented as liquid-containing tanks. As depicted in \Cref{fig:three_comp}, these tanks are named the aerobic energy source ($Ae$), anaerobic fast energy source ($AnF$) and anaerobic slow energy source ($AnS$). $Ae$ is assumed to have infinite volume, which is indicated by the fading color to the left. A pipe connects $Ae$ to the middle tank and has the maximal flow capacity $m^{Ae}$. The pipe from the right tank into the middle tank allows flow in both directions and the maximal flow capacity $m^{AnS}$ from $AnS$ into $AnF$ and $m^{AnF}$ from $AnF$ into $AnS$. A tap ($p$) is attached to the bottom of the middle tank $AnF$ and liquid flow from this tap represents energy demand. The fill levels of tanks and flows through pipes change as liquid flows from the tap. If the athlete expends energy, the liquid level of $AnF$ drops and initiates flow from $Ae$ and $AnS$ into $AnF$. When the middle tank is empty, liquid flow out of $p$ no longer matches the demand and exhaustion is assumed.

In the depicted situation in \Cref{fig:three_comp} $p$ was opened and liquid flowed out of $AnF$. As a result, the liquid level in $AnF$ dropped and thus liquid started to flow from $Ae$ into $AnF$. The more the fill level of the middle tank $AnF$ dropped, the less liquid pressured against the pipe exit from $Ae$, and the more the flow from $Ae$ increased. In \Cref{fig:three_comp} the flow out of $p$ was so large, that the fill level of $AnF$ dropped below the top of $AnS$ ($h>\theta$) and liquid from $AnS$ started to flow into $AnF$ too. The liquid volume in $AnF$ is limited such that it can only contribute to flow out of $p$ for a limited time. If the simulated athlete stopped exercise, then their power output would decrease to 0 and the tap $p$ would close. If the tap was closed in the depicted situation in \Cref{fig:three_comp}, liquid from the outer tanks $Ae$ and $AnS$ would refill the middle tank $AnF$ until its fill level rises above the fill level of $AnS$. Then, liquid from $Ae$ would continue to refill $AnF$ and liquid from $AnF$ would flow into $AnS$. The model would mimic recovery and $Ae$ would eventually refill both other tanks.

\citet{morton_three_1986} mathematically expressed liquid flows within this system as first- and second-order ordinary differential equations and in \citet{weigend_new_2021} we extended these equations so that they apply to all possible configurations of their generalized model. Due to liquid pressure and flow dynamics, varying fill levels of the middle and the right tank affect how flow from the tap $p$ is estimated. With these interactions between three tanks, \gls*{hyd-weig} is capable of predicting energy expenditure and recovery as a more complex function than the above introduced critical power and \gls*{w'bal-ode} models. 
It features eight adjustable parameters. Depicted in \Cref{fig:three_comp}, the parameters $\phi$, $\gamma$, $\theta$ represent tank positions, $AnF$, $AnS$ tank capacities, and $ m^{Ae}$, $m^{AnF}$, $m^{AnS}$ maximal flow capacities. A configuration $c$ for \gls*{hyd-weig} entails the positions, sizes and capacities of each tank and is therefore defined as 
\begin{equation} \label{eq:configuration}
    c = [ AnF, AnS, m^{Ae}, m^{AnS}, m^{AnF}, \phi, \theta, \gamma ].
\end{equation}
Fitting the \gls*{hyd-weig} to an athlete means finding the configuration that enables the model output to best reproduce the observed exercise responses of an athlete. In \citet{weigend_new_2021} we introduced an evolutionary computation workflow to derive such configurations. We fitted a configuration to \gls*{cp} and \gls*{w'} measures of an athlete as well as recovery ratios derived from a publication by~\cite{caen_reconstitution_2019}. The same recovery ratios are used for every fitting and thus, to fit \gls*{hyd-weig} to an athlete, only \gls*{cp} and \gls*{w'} of the athlete are required. These are the same measures that are needed to apply \gls*{w'bal-ode} models and, hence, the required input measures are the same for all compared \gls*{hyd-weig} and \gls*{w'bal-ode} models. With specific reference to our fitting method in~\citet{weigend_new_2021}, 10 evolutionary fittings to given \gls*{cp} and \gls*{w'} were estimated and the best fitting has been used.

\subsubsection{An additional \gls*{w'bal} model}
\label{subsec:add_wbal}

The above introduced models \gls*{w'bal-skib}, \gls*{w'bal-bart}, and \gls*{hyd-weig} were each created by fitting to different sets of recovery observations. Objective metrics to compare model quality, e.g. the Akaike Information Criterion \citep{burnham_multimodel_2004}, require models to be fitted to the same data. Therefore, to allow a more comprehensive comparison, we added a third \gls*{w'bal-ode} model with a new \gls*{tau-weig}. We derived this \gls*{tau-weig} with a procedure as close as possible to the ones of \citet{skiba_modeling_2012} and~\citet{bartram_accuracy_2018}.

As the first step, a constant value for $\Tau_t$ in \Cref{eq:differential_rec} was fitted to each recovery ratio and recovery time combination from Table 1 of the Appendix of \citet{weigend_new_2021}. For these observations power output was constant for every discrete time step $t$ during recovery and thus $\Tau_t$ could be considered as constant with the same value for all $t$. We used the standard Broyden-Fletcher-Goldfarb-Shanno algorithm implementation of SciPy \citep{scipy_10_contributors_scipy_2020} with 200 as the initial guess to fit a constant $\Tau_t$ that enabled \Cref{eq:differential_rec} to best reproduce the observed recovery ratio. This resulted in twelve pairs of fitted constant $\Tau_t$s to constant recovery intensities. 

As the next step, we then fitted an exponential function to these twelve pairs using the non-linear least squares implementation of SciPy \citep{scipy_10_contributors_scipy_2020}. With the recovery intensity as $D_{\gls*{cp}_t}$, the function was of the form \mbox{$\Tau_t = a \cdot e^{D_{\gls*{cp}_t}\cdot{b}} + c$}. With the values of \citet{skiba_modeling_2012} as the initial guess \mbox{(546, -0.01, 316)}, the resulting optimal constants were $a = 1274.45, b=-0.0308,$ and $ c= 266.65$. Thus, given any $D_{\gls{cp}_t}$ at a discrete time step $t$, $\Tau_{\mathrm{weig}_t}$ can be estimated as
\begin{equation}\label{eq:weigend_tau}
    \Tau_{\mathrm{weig}_t} = 1274.45 \cdot e^{-0.0308 \cdot D_{{\gls*{cp}}_t}} + 266.65.
\end{equation}
Unfortunately, this fitted equation failed to satisfactorily fit the data ($R^2=0.14$) but we nevertheless used it because it was developed using a procedure that closely resembled those used to estimate \gls*{tau-skib} and \gls{tau-bart}. The introduction of \gls*{tau-weig} is valuable because it allows the application of the Akaike Information Criterion metric, which requires compared models to be fitted to the same data points. Henceforth, \Cref{eq:differential_rec} with \Cref{eq:weigend_tau} will be referred to as \gls*{w'bal-weig}.

\subsection{Procedure for computing comparable recovery predictions}
\label{subsec:compare_procedure}

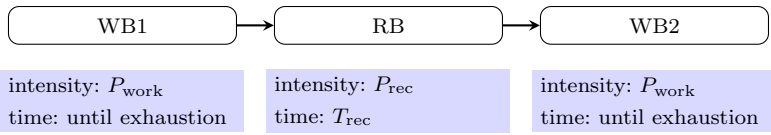
\begin{figure}
    \centering
    \begin{tikzpicture}[node distance=2cm]
        \node (WB1) [node] {WB1};
        \node (WB1desc) [below of=WB1, yshift=1cm, fill=blue!15, text width=3cm] {intensity: $P_\mathrm{work}$\\time: until exhaustion};
        \node (RB) [node, right of=WB1, xshift=1.5cm] {RB};
        \node (RBdesc) [below of=RB, yshift=1cm, fill=blue!15, text width=3cm] {intensity: $P_\mathrm{rec}$\\time: $T_\mathrm{rec}$};
        \node (WB2) [node, right of=RB, xshift=1.5cm] {WB2};
        \node (WB2desc) [below of=WB2, yshift=1cm, fill=blue!15, text width=3cm] {intensity: $P_\mathrm{work}$\\time: until exhaustion};
        \draw [arrow] (WB1) -- (RB);
        \draw [arrow] (RB) -- (WB2);
        \end{tikzpicture}
    \caption{A schematic of the protocol to estimate recovery ratios. An exhaustive work bout (WB1) at a set intensity ($P_\mathrm{work}$) is prescribed. Immediately after exhaustion is reached, a recovery bout (RB) follows at a lower recovery intensity ($P_\mathrm{rec}$) for a set duration ($T_\mathrm{rec}$). Then, a second exhaustive work bout (WB2) is conducted and the ratio of the time to exhaustion of WB2 to the one of WB1 represents the amount that was recovered during RB.}
    \label{fig:wb1_rb_wb2_protocol}
\end{figure}

We compare the abilities of above defined \gls*{w'bal-ode} and \gls*{hyd-weig} models to predict ``recovery ratios''. Recovery ratios are computed from exercise protocols involving two exhaustive work bouts (WB1 and WB2) interspersed with a recovery bout (RB). A schematic of the protocol is depicted in \Cref{fig:wb1_rb_wb2_protocol}. First, the model simulates exercise at a fixed work intensity ($P_\mathrm{work}$) above \gls*{cp} until exhaustion. Immediately after exhaustion is reached, exercise intensity switches to a lower recovery intensity ($P_\mathrm{rec}$) below \gls*{cp}. After a set time ($T_\mathrm{rec}$) at that recovery intensity, a second work bout (WB2) until exhaustion at $P_\mathrm{work}$ is simulated. The time to exhaustion of WB2 is expected to be shorter than the one of WB1 due to the limited recovery time in between the work bouts. Because it is assumed that \gls*{w'} is completely depleted at the end of WB1, the ratio of the second time to exhaustion to the first represents the amount of \gls*{w'} recovered. Thus, the time to exhaustion of WB2 divided by the time to exhaustion of WB1 multiplied by 100 results in a recovery ratio in percent (\%). 

This outlined procedure aligns with the assumptions of \gls*{cp} or \gls*{w'bal-ode} models and enables the direct comparison of the simulated recovery ratios from each model \gls*{hyd-weig}, \gls*{w'bal-skib}, \gls*{w'bal-bart}, \gls*{w'bal-weig}, and with the published data. As an example, the recovery ratio curves in \Cref{fig:caen_2021_rec} were obtained by estimating the WB1 $\rightarrow$ RB $\rightarrow$ WB2 protocol for every recovery duration ($T_\mathrm{rec}$) in seconds between 0 and 900 seconds. Published observations by \citet{caen_w_2021} were added to the plot.

\subsection{Data extraction for model comparisons}
\label{subsec:comparison_studies}

We extracted data from previous studies that investigated energy recovery dynamics and used to it compare and evaluate recovery ratio predictions of all models. The studies for comparison were identified from Table 1 of the comprehensive review by~\citet{chorley_application_2020}. From these studies, we retained those that featured appropriate data, except those that met the following exclusion criteria:
\begin{itemize}
    \item Featured a mode of exercise other than cycling. Cycle ergometers measure power output directly. Power outputs during modes of exercise like running or swimming are not directly comparable because they are estimated using different methods or are approximated, e.g, \citep{morton_critical_2004} focused only on speed instead of power. 
    \item The observations were made under extreme conditions, e.g., hypoxia or altitude.
    \item Insufficient information was reported to simulate the prescribed protocol in and/or to infer a recovery ratio of \gls*{w'} in percent, e.g., the integral version of the \gls*{w'bal} model by \citet{skiba_modeling_2012} assumes recovery during high-intensity exercise such that recovery ratios cannot be straightforwardly inferred.
    \item The prescribed protocol leaves doubt if reported recovery ratios are comparable to the ``recovery estimation protocol'' described earlier, e.g., repeated ramp tests until exhaustion, 50\% \gls*{w'} depletion followed by a 3-min all-out test, or knee-extension maximal voluntary contraction (MVC) test during recovery.
\end{itemize}
Five studies were included for comparison, four of which were obtained from the Chorley and Lamb review: \citep{bartram_accuracy_2018}, \citep{chidnok_exercise_2012}, \citep{ferguson_effect_2010}, and \citep{caen_reconstitution_2019}. After the summary of \citet{chorley_application_2020} was published, \citet{caen_w_2021} published a study that investigated the \gls*{w'} reconstitution dynamics in even more detail and which was thus added to the list. 

The data in the listed studies were presented in diverse ways, such that modifications were made to some of the data to enable model comparison. The study by \citet{caen_reconstitution_2019} did not report distinct mean values for every investigated condition, such that we derived approximate values in \citet{weigend_new_2021} to fit our \gls*{hyd-weig} to their conditions. Hence, the data for comparison are the values from~\citet{weigend_new_2021}. Further, the study by \citet{bartram_accuracy_2018} fitted their own \gls*{w'bal-bart} model, where $\Tau_t$ is defined according to \Cref{eq:bartram_tau}. Therefore, we used \gls*{w'bal-bart} model predictions for prescribed intensities of \citet{bartram_accuracy_2018} as the observations against which the other models were compared.

The study by \citet{chidnok_exercise_2012} reported times to exhaustion from their intermittent exercise protocol instead of recovery ratios. Power output during recovery was constant in their tests. Therefore, in order to derive recovery ratio estimations that are comparable with the \mbox{WB1 $\rightarrow$ RB $\rightarrow$ WB2} procedure defined above, we fitted a constant value for $\Tau_t$ of the \gls*{w'bal-ode} model to each of their prescribed protocols and times to exhaustion. These constant values for $\Tau_t$ were fitted with the Brent method implementation by SciPy \citep{scipy_10_contributors_scipy_2020} to find a local minimum in the interval between [100, 1000]. We then used \mbox{WB1 $\rightarrow$ RB $\rightarrow$ WB2} recovery ratio estimations of \gls*{w'bal-ode} models with fitted constant $\Tau_t$ as the observations with which to compare \gls*{w'bal-skib}, \gls*{w'bal-bart}, \gls*{w'bal-weig}, and \gls*{hyd-weig}.

\subsection{The metrics of goodness of fit}
\label{subsec:metrics_of_fit}

The metrics of goodness of fit used to compare the models were \gls*{RMSE}, \gls*{MAE}, and \gls*{AIC}. \Citet{chai_root_2014} discussed \gls*{RMSE} and \gls*{MAE} as widely adopted metrics for assessing model prediction capabilities. We compared predictive accuracy by comparing \gls*{RMSE} and \gls*{MAE} on data to which competing models were not fitted. Lower values for \gls*{RMSE} and \gls*{MAE} were interpreted as more accurate predictions. 

To statistically compare prediction error distributions between models, we used a bootstrap hypothesis test \citep{efron_introduction_1993,good_permutation_2000}. We did so because only small data sets were available and we could not assume normal distributed prediction errors with equal variances for every compared model. The null hypothesis of our bootstrap test was that prediction error distributions of two compared models are the same. Because we used two prediction error metrics (\gls*{RMSE} and \gls*{MAE}) we investigated the null hypothesis on both. We used the absolute difference between RMSE and also between MAE of compared groups as our test statistics. With the null hypothesis that error distributions are the same, we could bootstrap new samples by randomly selecting with replacement from all pooled observations. We created a distribution of test statistics from 1\;000\;000 bootstrap samples to reliably approximate the p-value of our observed test statistic at high precision. We rejected the null hypothesis if the p-value$<.05$.

We also compared models with the \gls*{AIC}, which was first proposed by \cite{sugiura_further_1978}. The \gls*{AIC} is a model selection tool to investigate the balance between model complexity and explanatory capability \citep{burnham_multimodel_2004}. \gls*{AIC} penalizes the number of parameters of the model and thus provides insight into the balance between model complexity and goodness of fit. The lower the \gls*{AIC} score, the better this balance is met. The \gls{AIC} was calculated as
\begin{equation}\label{eq:aic}
    \text{AIC}_c = n \cdot \text{ln}(\text{MSE}) + 2k + \frac{2k \cdot (k+1)}{n-k-1},
\end{equation}
where MSE is the Mean Squared Error, $n$ is the number of data points and $k$ is the number of parameters of the model. Models have to be fitted to and applied to the same data in order to obtain comparable \gls*{AIC} scores. Therefore, only \gls*{w'bal-weig} and \gls*{hyd-weig} were comparable with this criterion in this work.

Altogether, the hypothesis that the more complex \gls*{hyd-weig} model fits the data better than the established \gls*{w'bal-ode} models will be supported if the overall \gls*{RMSE}, \gls*{MAE} and \gls*{AIC} scores are lower for \gls*{hyd-weig} than for other models, and if prediction error distributions are significantly different to those of other models.

\section{Results}
\label{sec:results}

In the following section, we present the extracted data and the prediction results of \gls*{w'bal-ode} and \gls*{hyd-weig} models for each listed previous study. We refer to extracted data from studies by the last name of the first author, e.g., the extracted data from \citet{bartram_accuracy_2018} is referred to as ``Bartram data set''. All studies collected their data through performance tests that required athletes to exercise until volitional exhaustion. Such tests are affected by circumstances that are hard to measure and control, e.g, motivation, nutrition, and state-of-mind. Therefore, recovery ratio observations are noisy and the extracted group averages were accompanied by large standard deviations. These uncertainties prevented us from drawing conclusions about model quality on averages of individual data sets and instead necessitated to perform the comparison between models across all available data. We begin by presenting the extracted data and model predictions of individual data sets throughout \Cref{subsec:bart_data,subsec:caen_data,subsec:chidnok_results,subsec:ferg_data,subsec:weig_data} followed by summarizing all prediction errors and resulting \gls*{RMSE}, \gls*{MAE}, and \gls*{AIC} scores in the final \Cref{subsec:metrics_summary}.

\subsection{Bartram data set}
\label{subsec:bart_data}
\begin{figure}
    \centering
    \includegraphics[width=\linewidth]{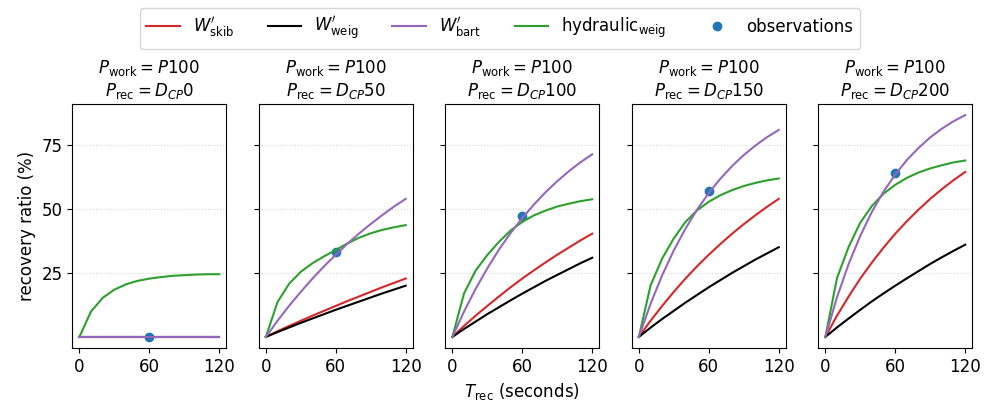}
    \caption{Comparison of model predictions with the defined recovery estimation protocol \mbox{(WB1 $\rightarrow$ RB $\rightarrow$ WB2)}. Depicted are the recovery dynamics around 60 seconds at various $D_{CP}$ recovery intensities after a preceding exhaustive exercise at the intensity that is predicted to lead to exhaustion after 100 seconds ($P100$). Chosen intensities and time frames stem from the protocol prescribed by~\citet{bartram_accuracy_2018} and predictions of \gls*{w'bal-bart} were used as the observations with which to compare models.}
    \label{fig:bartram_study}
\end{figure}
\begin{table}
\begin{adjustwidth}{-.1in}{-.1in}\centering 
{\footnotesize 
\caption{The left part of the table summarizes extracted data and conditions from \citet{bartram_accuracy_2018}. The right part of the table displays model predictions. Predictions of $W^\prime_\mathrm{bart}$ were taken as the observed recovery ratios.} 
\label{tab:bart_comp} 
\begin{tabular}{ 
S[table-format=3.0] 
S[table-format=5.0] 
S[table-format=3.0] 
S[table-format=3.0] 
S[table-format=3.0] 
S[table-format=3.1] 
S[table-format=3.1]
S[table-format=3.1]
S[table-format=3.1]
}
\toprule\multicolumn{5}{c}{Parameters from \citet{bartram_accuracy_2018}}&
\multicolumn{1}{c}{Observed}&
\multicolumn{3}{c}{Predicted recovery ratio} \\
\cmidrule{1-5} \cmidrule{7-9}
\multicolumn{1}{c}{\gls*{cp}}&
\multicolumn{1}{c}{\gls*{w'}}&
\multicolumn{1}{c}{$P_\mathrm{work}$}&
\multicolumn{1}{c}{$P_\mathrm{rec}$}&
\multicolumn{1}{c}{$T_\mathrm{rec}$}&
\multicolumn{1}{c}{recovery ratio}
&\multicolumn{1}{c}{$W^\prime_\mathrm{skib}$}&\multicolumn{1}{c}{$W^\prime_\mathrm{weig}$}&\multicolumn{1}{c}{$\mathrm{hydraulic}_\mathrm{weig}$}\\
 \midrule
393&23300&626&393&60&0.0&0.0&0.0&22.7\\
393&23300&626&343&60&33.0&12.1&10.6&33.9\\
393&23300&626&293&60&47.0&22.8&16.9&44.8\\
393&23300&626&243&60&57.0&32.1&19.4&52.7\\
393&23300&626&193&60&64.0&40.3&20.0&59.3\\
\bottomrule 
\end{tabular}
}\end{adjustwidth} 
\end{table}
The protocol prescribed by \citet{bartram_accuracy_2018} consisted of three work bouts interspersed with two 60-second recovery bouts. The first two work bouts each lasted for 30 seconds, and the final one until volitional exhaustion. Work bout exercise intensity ($P_\mathrm{work}$) was set to $P100$, i.e., the intensity that was predicted to lead to exhaustion after 100 seconds. The recovery bout intensity ($P_\mathrm{rec}$) was set to differences to \gls*{cp} ($D_{CP}$) of 200, i.e., \gls*{cp} - 200 watts, or 150, 100, 50, or 0. The group averaged CP and W' for the four world-class cyclists featured in \citet{bartram_accuracy_2018} were 393 watts and 23,300 joules. Altogether, these input values resulted in an estimated $P100$ exhaustive intensity of 626 watts and recovery intensities $D_{CP}$ 0 of 393 watts, $D_{CP}$ 50 of 343 watts, $D_{CP}$ 100 of 293 watts, $D_{CP}$ 150 of 243 watts, and $D_{CP}$ 200 of 193 watts, respectively. 

The resulting recovery predictions of \gls*{w'bal-ode} and \gls*{hyd-weig} models are summarized in \Cref{fig:bartram_study} and \Cref{tab:bart_comp}. The \gls*{w'bal-bart} model was not compared because it was the model that \citet{bartram_accuracy_2018} fitted to their observations and we used it to create the observations against which the other models were compared. The fitted \gls*{hyd-weig} configuration to \gls*{cp} and \gls*{w'} by~\citet{bartram_accuracy_2018} was: $[$23111.91, 65845.28, 391.57, 148.88, 24.15,  0.73, 0.01, 0.24$]$. 

\Cref{fig:bartram_study} and \Cref{tab:bart_comp} show that in all cases except $D_{CP}$0 the recovery ratios predicted by \gls*{hyd-weig} model were closest to the ones observed by \citet{bartram_accuracy_2018}, followed by \gls*{w'bal-skib} and then \gls*{w'bal-weig}. On the contrary, the hydraulic model is the only model to predict recovery at $D_{CP}$0.

\subsection{Caen data set}
\label{subsec:caen_data}

\begin{figure}[b]
    \begin{center}
        \includegraphics[width=0.8\linewidth]{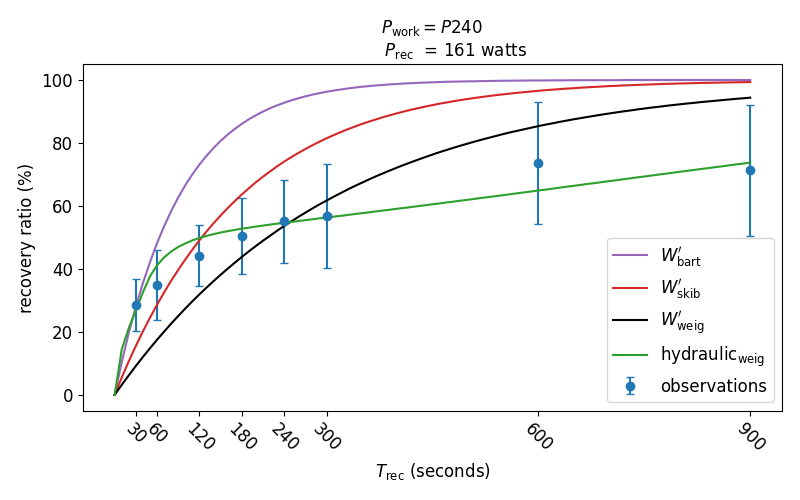}
        \caption{Comparison of model predictions with published observations by~\citet{caen_w_2021}. After an exhaustive exercise bout at $P240$, recovery dynamics at an intensity of 161 watts were simulated using the defined recovery estimation protocol of this work \mbox{(WB1 $\rightarrow$ RB $\rightarrow$ WB2)}. Published observed recovery ratios by \citet{caen_w_2021} are depicted in blue.}
        \label{fig:caen_2021_rec}
    \end{center}
\end{figure}

\begin{table} [b]
\begin{adjustwidth}{-.3in}{-.3in}\centering 
{\footnotesize 
\caption{The left part of the table summarizes extracted data and conditions from \citet{caen_w_2021}. The right part of the table displays model predictions.} 
\label{tab:caen_comp} 
\begin{tabular}{ 
S[table-format=3.0] 
S[table-format=5.0] 
S[table-format=3.0] 
S[table-format=3.0] 
S[table-format=3.0] 
S[table-format=3.1] 
S[table-format=3.1]
S[table-format=3.1]
S[table-format=3.1]
S[table-format=3.1]
}
\toprule\multicolumn{5}{c}{Parameters from \citet{caen_w_2021}}&
\multicolumn{1}{c}{Observed}&
\multicolumn{4}{c}{Predicted recovery ratio} \\
\cmidrule{1-5} \cmidrule{7-10}
\multicolumn{1}{c}{\gls*{cp}}&
\multicolumn{1}{c}{\gls*{w'}}&
\multicolumn{1}{c}{$P_\mathrm{work}$}&
\multicolumn{1}{c}{$P_\mathrm{rec}$}&
\multicolumn{1}{c}{$T_\mathrm{rec}$}&
\multicolumn{1}{c}{recovery ratio}
&\multicolumn{1}{c}{$W^\prime_\mathrm{bart}$}&\multicolumn{1}{c}{$W^\prime_\mathrm{skib}$}&\multicolumn{1}{c}{$W^\prime_\mathrm{weig}$}&\multicolumn{1}{c}{$\mathrm{hydraulic}_\mathrm{weig}$}\\
 \midrule
269&19200&349&161&30&28.6&28.0&15.5&9.2&26.9\\
269&19200&349&161&60&34.8&48.2&28.7&17.5&41.2\\
269&19200&349&161&120&44.2&73.2&49.1&31.9&49.8\\
269&19200&349&161&180&50.5&86.1&63.7&43.8&52.8\\
269&19200&349&161&240&55.1&92.8&74.1&53.6&54.7\\
269&19200&349&161&300&56.8&96.3&81.5&61.8&56.3\\
269&19200&349&161&600&73.7&99.9&96.6&85.4&64.9\\
269&19200&349&161&900&71.3&100.0&99.4&94.4&73.8\\
\bottomrule 
\end{tabular}
}\end{adjustwidth} 
\end{table}

The protocol by \citet{caen_w_2021} investigated the recovery dynamics following exhaustive exercise at $P240$ (published average of 349 watts). They prescribed a recovery intensity of 161 watts on average, which was determined by selecting $90\%$ of the power at gas exchange threshold \citep{binder_methodological_2008} of their participants. The average \gls*{cp} of their participants was 269 watts, and the average \gls*{w'} 19\,200 joules. The reported observed recovery ratios were $28.6\pm8.2\%$ after 30 seconds, $34.8\pm11.1\%$ after 60 seconds, $44.2\pm9.7\%$ after 120 seconds, $50.5\pm12.1\%$ after 180 seconds, $55.1\pm13.3\%$ after 240 seconds, $56.8\pm16.4\%$ after 300 seconds, $73.7\pm19.3\%$ after 600 seconds, and $71.3\pm20.8\%$ after 900 seconds.

The simulation parameters and results of the defined recovery estimation protocol are summarized in \Cref{fig:caen_2021_rec} and \Cref{tab:caen_comp}. Fitting \gls*{hyd-weig} to \gls*{cp} and \gls*{w'} group averages resulted in the configuration $[$17631.06, 46246.13, 267.28, 117.50, 20.09, 0.68, 0.01, 0.29$]$. The recovery ratios predicted by the \gls*{hyd-weig} model better matched the observed values compared to all the other models. Nevertheless, some lack of fit for the \gls{hyd-weig} model was observed: the model overpredicted the recovery ratios at early time points and underpredicted those at longer time points except for the last one. \gls*{w'bal-skib}, \gls*{w'bal-bart}, and \gls*{w'bal-weig} model predictions consistently overestimated recovery for longer recovery times. 

\subsection{Chidnok data set}
\label{subsec:chidnok_results}

\cite{chidnok_exercise_2012} prescribed a protocol that alternated between 60-second work bouts and 30-second recovery bouts until the athlete reached exhaustion. With their protocol, the work-bout intensity $P_\mathrm{work}$ was set to $P240$. The protocol prescribed four trials each with a different recovery intensity $P_\mathrm{rec}$ (20 watts as the ``low'' recovery intensity, 95 watts as ``medium'', 173 watts as ``high'', and 270 watts as the ``severe'' recovery intensity). The participants had an average \gls*{cp} of 241 watts and \gls*{w'} of 21\,100 joules. Their recorded times to exhaustion were $1224\pm497$ seconds with ``low'' recovery intensity, $759\pm243$ seconds with ``medium'', $557\pm90$ seconds with ``high'', and $329\pm29$ seconds with ``severe''. 

\begin{figure}[b]
    \centering
    \includegraphics[width=\linewidth]{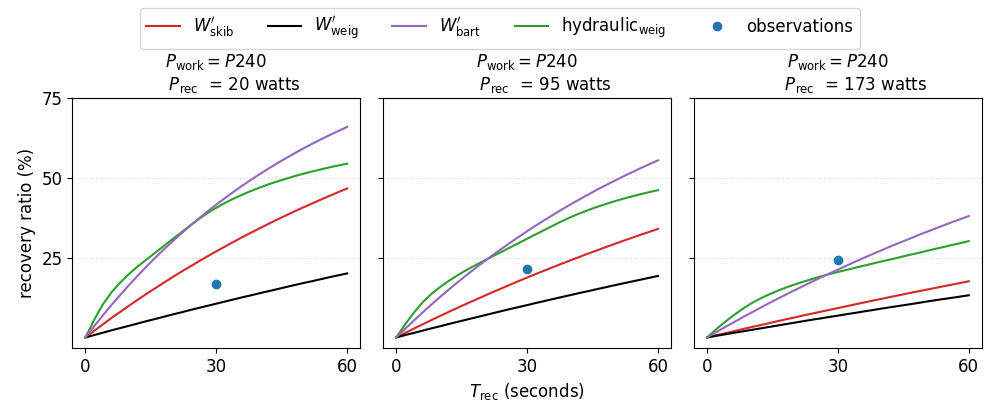}
    \caption{Predicted recovery dynamics of compared models up to 60 seconds after a preceding exhaustive exercise at $P240$ and at three different recovery intensities (20 watts, 95 watts, and 173 watts). Observations were predicted recovery ratios of \gls*{w'bal-ode} models with a constant $\Tau$ fitted to reported times to exhaustion by \citet{chidnok_exercise_2012}.}
    \label{fig:chidnok_study}
\end{figure}
\begin{table}
\begin{adjustwidth}{-.3in}{-.3in}\centering 
{\footnotesize 
\caption{The left part of the table summarizes extracted data and conditions from \citet{chidnok_exercise_2012}. The right part of the table displays model predictions.} 
\label{tab:chid_comp} 
\begin{tabular}{ 
S[table-format=3.0] 
S[table-format=5.0] 
S[table-format=3.0] 
S[table-format=3.0] 
S[table-format=3.0] 
S[table-format=3.1] 
S[table-format=3.1]
S[table-format=3.1]
S[table-format=3.1]
S[table-format=3.1]
}
\toprule\multicolumn{5}{c}{Parameters from \citet{chidnok_exercise_2012}}&
\multicolumn{1}{c}{Observed}&
\multicolumn{4}{c}{Predicted recovery ratio} \\
\cmidrule{1-5} \cmidrule{7-10}
\multicolumn{1}{c}{\gls*{cp}}&
\multicolumn{1}{c}{\gls*{w'}}&
\multicolumn{1}{c}{$P_\mathrm{work}$}&
\multicolumn{1}{c}{$P_\mathrm{rec}$}&
\multicolumn{1}{c}{$T_\mathrm{rec}$}&
\multicolumn{1}{c}{recovery ratio}
&\multicolumn{1}{c}{$W^\prime_\mathrm{bart}$}&\multicolumn{1}{c}{$W^\prime_\mathrm{skib}$}&\multicolumn{1}{c}{$W^\prime_\mathrm{weig}$}&\multicolumn{1}{c}{$\mathrm{hydraulic}_\mathrm{weig}$}\\
 \midrule
241&21100&329&20&30&16.6&41.6&27.0&10.6&40.6\\
241&21100&329&95&30&21.4&33.3&18.8&10.1&30.9\\
241&21100&329&173&30&24.4&21.3&9.2&6.8&20.5\\
\bottomrule 
\end{tabular}
}\end{adjustwidth} 
\end{table}

As described in \Cref{subsec:comparison_studies}, in order to compare observations of~\citet{chidnok_exercise_2012} to \mbox{WB1 $\rightarrow$ RB $\rightarrow$ WB2} protocol estimations, a constant value for $\Tau_t$ for the \gls*{w'bal-ode} model was fitted to the protocol by \citet{chidnok_exercise_2012} for each of their recovery conditions. The resulting $\Tau_t$ values were 165.19 seconds for the ``low'' recovery intensity protocol, a $\Tau_t$ of 124.81 seconds for ``medium'', and a $\Tau_t$ of 107.45 seconds for ``high''. The ``severe'' recovery intensity was left out because 270 watts lies above the average \gls*{cp} of 241 watts. In this case, no recovery should occur if the assumptions of \gls*{w'bal-ode} model hold true. \citet{chidnok_exercise_2012} prescribed recovery bouts of 30 seconds and \mbox{WB1 $\rightarrow$ RB $\rightarrow$ WB2} protocol estimations with corresponding fitted $\Tau_t$s and with a $T_\mathrm{rec}$ of 30 seconds were  24.6\% at the ``low'' intensity, 21.7\% at ``medium'', and 16.7\% at the ``high'' recovery intensity.

The fitted \gls*{hyd-weig} configuration to \gls*{cp} and \gls*{w'} by~\citet{chidnok_exercise_2012} was $[$18919.76, 48051.77, 239.55, 115.05, 19.48, 0.68, 0.05, 0.31$]$. Predictions of all models and extracted conditions for the recovery estimation protocol are summarized in \Cref{fig:chidnok_study} and \Cref{tab:chid_comp}. In the case of the ``low'' recovery intensity predictions of the \gls*{w'bal-skib} and \gls*{w'bal-weig} models were the most accurate. In the case of the ``medium'' recovery intensity the \gls*{w'bal-skib} model was the most accurate, and in the remaining ``high'' condition \gls*{hyd-weig} and \gls*{w'bal-bart} model predictions were closest to the data. None of the models made predictions that were close to all three of the observations.

\subsection{Ferguson data set}
\label{subsec:ferg_data}

\begin{figure}
    \centering
    \includegraphics[width=0.8\linewidth]{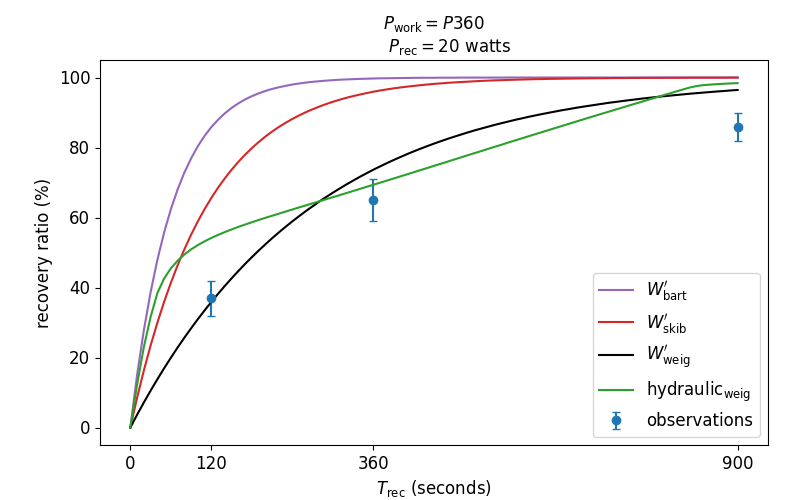}
    \caption{A comparison of predicted recovery dynamics after an exhaustive exercise bout at $P360$ and at a recovery intensity of 20 watts. Recovery ratios are estimated with the \mbox{(WB1 $\rightarrow$ RB $\rightarrow$ WB2)} protocol, which resembles the prescribed protocol by \citet{ferguson_effect_2010}. Published observations by \citet{ferguson_effect_2010} are depicted in blue.}
    \label{fig:ferguson_study}
\end{figure}

\citet{ferguson_effect_2010} prescribed  a protocol with an initial time to exhaustion bout at the intensity that was predicted to lead to exhaustion after 360 seconds ($P360$) followed by a recovery at 20 watts for 2 minutes, 6 minutes, or 15 minutes. After recovery, exercise intensity was then increased back to one of three possible high-intensity work rates. Thus, each participant performed nine tests in total with three different constant work rates after three different recovery times. The \gls*{cp} model was fitted to these three times to exhaustion after each recovery period to determine changes in \gls*{cp} and \gls*{w'}. \citet{ferguson_effect_2010} published their group averages for \gls*{cp} as 212 watts, \gls*{w'} as 21\,600 joules, the $P360$ as 269 watts, and the observed recovery ratios after 2 minutes as ($37\% \pm 5\%$), 6 minutes ($65\% \pm 6\%$), and 15 minutes ($86\% \pm 4\%$).

\begin{table}
\begin{adjustwidth}{-.3in}{-.3in}\centering 
{\footnotesize 
\caption{The left part of the table summarizes extracted data and conditions from \citet{ferguson_effect_2010}. The right part of the table displays model predictions.} 
\label{tab:ferg_comp} 
\begin{tabular}{ 
S[table-format=3.0] 
S[table-format=5.0] 
S[table-format=3.0] 
S[table-format=3.0] 
S[table-format=3.0] 
S[table-format=3.1] 
S[table-format=3.1]
S[table-format=3.1]
S[table-format=3.1]
S[table-format=3.1]
}
\toprule\multicolumn{5}{c}{Parameters from \citet{ferguson_effect_2010}}&
\multicolumn{1}{c}{Observed}&
\multicolumn{4}{c}{Predicted recovery ratio} \\
\cmidrule{1-5} \cmidrule{7-10}
\multicolumn{1}{c}{\gls*{cp}}&
\multicolumn{1}{c}{\gls*{w'}}&
\multicolumn{1}{c}{$P_\mathrm{work}$}&
\multicolumn{1}{c}{$P_\mathrm{rec}$}&
\multicolumn{1}{c}{$T_\mathrm{rec}$}&
\multicolumn{1}{c}{recovery ratio}
&\multicolumn{1}{c}{$W^\prime_\mathrm{bart}$}&\multicolumn{1}{c}{$W^\prime_\mathrm{skib}$}&\multicolumn{1}{c}{$W^\prime_\mathrm{weig}$}&\multicolumn{1}{c}{$\mathrm{hydraulic}_\mathrm{weig}$}\\
 \midrule
212&21600&269&20&120&37.0&85.8&65.6&35.9&54.2\\
212&21600&269&20&360&65.0&99.7&95.9&73.6&69.4\\
212&21600&269&20&900&86.0&100.0&100.0&96.4&98.4\\
\bottomrule 
\end{tabular}
}\end{adjustwidth} 
\end{table}

Extracted parameters for the recovery intensity protocol and model prediction results are summarized in \Cref{fig:ferguson_study} and \Cref{tab:chid_comp} together with reported means by~\cite{ferguson_effect_2010}. The fitted \gls*{hyd-weig} configuration to \gls*{cp} and \gls*{w'} group averages by~\citet{ferguson_effect_2010} was $[$18730.05, 81030.54, 211.56, 94.31, 18.76, 0.63, 0.21, 0.34$]$. In this setup \gls*{w'bal-weig} was overall closest to published observations. \Gls*{hyd-weig} overestimated the recovery after 120 and after 900 seconds. \Gls*{w'bal-skib} and \gls*{w'bal-bart} overestimated recovery in every instance.

\subsection{Weigend data set}
\label{subsec:weig_data}

We derived the values from Table 1 in the Appendix of our \citet{weigend_new_2021} publication from~\citet{caen_reconstitution_2019}. Reported measures recreate the depicted means in Figure 3 of the publication by~\citet{caen_reconstitution_2019}. They consisted of three recovery ratios for four conditions each: Preceding exhausting exercise at $P240$ or $P480$ followed by recovery at 33\% of \gls*{cp} or 66\% of \gls*{cp}. The participants of~\citet{caen_reconstitution_2019} had an average \gls*{cp} of 248 watts and \gls*{w'} of 18\,200 joules, which results in a $P240$ of 285 watts, a $P480$ of 323 watts, 33\% of \gls*{cp} as 81 watts, and 66\% of \gls*{cp} as 163 watts.

\begin{figure}
    \centering
    \includegraphics[width=0.9\linewidth]{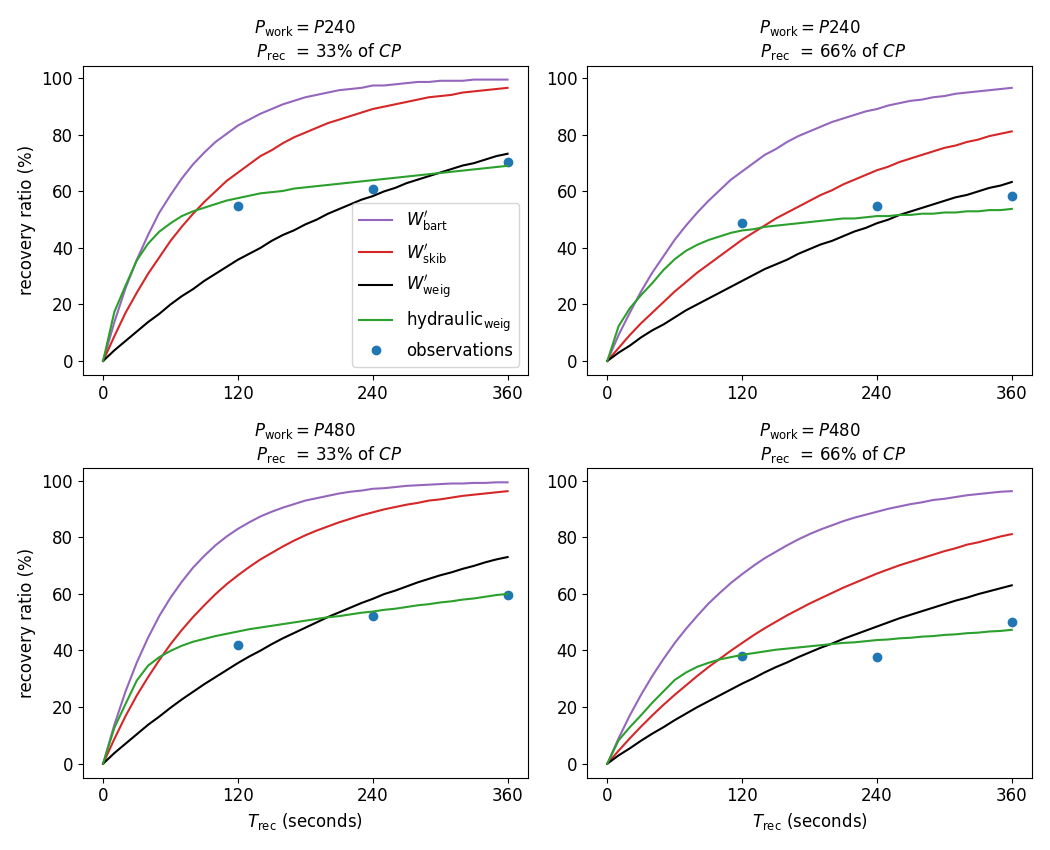}
    \caption{Predicted recovery dynamics in comparison to measures that we derived from observations of \cite{caen_reconstitution_2019}. We derived three recovery ratios for four conditions each: Preceding exhausting exercise at $P240$ or $P480$ followed by recovery at 33\% of \gls*{cp} or 66\% of \gls*{cp}. Depicted observations are the values from Table 1 in the Appendix of our publication~\citet{weigend_new_2021} and approximate Figure 3 of the publication by~\citet{caen_reconstitution_2019}. \gls*{w'bal-weig} and \gls*{hyd-weig} were fitted to these observations.}
    \label{fig:weig_study}
\end{figure}

\begin{table} 
\begin{adjustwidth}{-.3in}{-.3in}\centering 
{\footnotesize 
\caption{The left part of the table summarizes extracted data and conditions that \citet{weigend_new_2021} derived from \citet{caen_reconstitution_2019}. The right part of the table displays model predictions. Both $W^\prime_\mathrm{weig}$ and $\mathrm{hydraulic}_\mathrm{weig}$ were fitted to these observations. Their predicted recovery ratios were recorded for the \gls*{AIC} goodness of fit estimation metric in \Cref{subsec:metrics_of_fit}.} 
\label{tab:weigend_comp} 
\begin{tabular}{ 
S[table-format=3.0] 
S[table-format=5.0] 
S[table-format=3.0] 
S[table-format=3.0] 
S[table-format=3.0] 
S[table-format=3.1] 
S[table-format=3.1]
S[table-format=3.1]
S[table-format=3.1]
S[table-format=3.1]
}
\toprule\multicolumn{5}{c}{Parameters from \citet{weigend_new_2021}}&
\multicolumn{1}{c}{Observed}&
\multicolumn{4}{c}{Predicted recovery ratio} \\
\cmidrule{1-5} \cmidrule{7-10}
\multicolumn{1}{c}{\gls*{cp}}&
\multicolumn{1}{c}{\gls*{w'}}&
\multicolumn{1}{c}{$P_\mathrm{work}$}&
\multicolumn{1}{c}{$P_\mathrm{rec}$}&
\multicolumn{1}{c}{$T_\mathrm{rec}$}&
\multicolumn{1}{c}{recovery ratio}
&\multicolumn{1}{c}{$W^\prime_\mathrm{bart}$}&\multicolumn{1}{c}{$W^\prime_\mathrm{skib}$}&\multicolumn{1}{c}{$W^\prime_\mathrm{weig}$}&\multicolumn{1}{c}{$\mathrm{hydraulic}_\mathrm{weig}$}\\
 \midrule
248&18200&323&81&120&55.0&83.1&66.7&35.5&58.3\\
248&18200&323&81&240&61.0&97.1&89.0&58.3&65.0\\
248&18200&323&81&360&70.5&99.5&96.3&73.1&70.1\\
248&18200&323&163&120&49.0&67.2&42.9&28.4&46.5\\
248&18200&323&163&240&55.0&89.2&67.4&48.7&51.5\\
248&18200&323&163&360&58.0&96.5&81.4&63.3&54.2\\
248&18200&285&81&120&42.0&83.0&66.8&35.5&46.8\\
248&18200&285&81&240&52.0&97.1&89.0&58.3&54.0\\
248&18200&285&81&360&59.5&99.5&96.3&73.1&60.4\\
248&18200&285&163&120&38.0&67.2&42.9&28.4&38.5\\
248&18200&285&163&240&37.5&89.3&67.4&48.7&43.6\\
248&18200&285&163&360&50.0&96.5&81.4&63.3&47.4\\
\bottomrule 
\end{tabular}
}\end{adjustwidth} 
\end{table}

Extracted parameters for the recovery ratio estimation protocol and model predictions are summarized in \Cref{fig:weig_study} and \Cref{tab:weigend_comp}. The best fit \gls*{hyd-weig} configuration to \gls*{cp} and \gls*{w'} group averages was $[$18042.06, 46718.18, 247.4, 106.77, 16.96, 0.72, 0.02, 0.25$]$. As described earlier, the recovery ratio values by \citet{weigend_new_2021} were used to fit the \gls*{tau-weig} for the \gls*{w'bal-weig} model and they are used in the evolutionary fitting process for \gls*{hyd-weig} to fit recovery dynamics. Therefore, both \gls*{w'bal-weig} and \gls*{hyd-weig} were not scrutinized for predictive accuracy on this data set. Their predicted recovery ratios were recorded for the \gls*{AIC} goodness of fit estimation metric in covered in the next subsection. Out of the remaining two models predictions of \gls*{w'bal-skib} were closer to the observations but both overpredict in nearly all instances.

\subsection{Summary of metrics of goodness of fit}
\label{subsec:metrics_summary}

\begin{table} 
\begin{adjustwidth}{-.3in}{-.3in}\centering 
\footnotesize 
\caption{Summary of the model prediction errors and estimated metric scores. The first two columns summarize prediction errors used to compare predictive accuracy of \gls*{w'bal-bart} and \gls*{hyd-weig} via \gls*{MAE}, standard deviation of absolute errors (SD), and \gls*{RMSE}. Prediction accuracy had to be be assessed using data to which models were not fitted to, such that we had to exclude the Bartram and Weigend data sets. The subsequent three columns summarize the prediction errors used to compare the predictive accuracies of \gls*{w'bal-skib}, \gls*{w'bal-weig} and \gls*{hyd-weig}. Here, the Weigend data set was excluded because \gls*{w'bal-weig} and \gls*{hyd-weig} were fitted to it. Finally, the \gls*{AIC} metric requires models to be fitted to and to be evaluated on the same data. Therefore, we compared \gls*{AIC} scores estimated from prediction errors of \gls*{w'bal-weig} and \gls*{hyd-weig} on all data sets. We approximated p-values for absolute differences in \gls*{MAE} and \gls*{RMSE} with a bootstrap hypothesis tests and considered $\mathrm{p}<.05$ as significant. For every metric, a lower score means a better result.}\label{tab:error_summary} 
\begin{threeparttable} 
\begin{tabular}{ 
c c 
S[table-format=4.2] 
S[table-format=4.2] 
S[table-format=4.2] 
S[table-format=4.2] 
S[table-format=4.2] 
S[table-format=4.2] 
S[table-format=4.2]} 
\toprule 
\multicolumn{2}{c}{} & 
\multicolumn{2}{c}{data for prediction scores 1} & 
\multicolumn{3}{c}{data for prediction scores 2} & 
\multicolumn{2}{c}{data for \gls*{AIC} scores} \\ 
\cmidrule(r){3-4} 
\cmidrule(r){5-7} 
\cmidrule(r){8-9} 
\multicolumn{2}{c}{} & \multicolumn{1}{c}{\gls*{w'bal-bart}} & \multicolumn{1}{c}{\gls*{hyd-weig}} & \multicolumn{1}{c}{\gls*{w'bal-skib}} & \multicolumn{1}{c}{\gls*{w'bal-weig}} & \multicolumn{1}{c}{\gls*{hyd-weig}} & \multicolumn{1}{c}{\gls*{w'bal-weig}} & \multicolumn{1}{c}{\gls*{hyd-weig}} \\ 
\midrule 
\multirow{5}{*}{\rotatebox[origin=c]{90}{Bartram}} 
 & 0 &  &  & 0.0 & 0.0 & 22.7 & 0.0 & 22.7 \\ 
 & 1 &  &  & -20.9 & -22.4 & 0.9 & -22.4 & 0.9 \\ 
 & 2 &  &  & -24.2 & -30.1 & -2.2 & -30.1 & -2.2 \\ 
 & 3 &  &  & -24.9 & -37.6 & -4.3 & -37.6 & -4.3 \\ 
 & 4 &  &  & -23.7 & -44.0 & -4.7 & -44.0 & -4.7 \\ 
\\ 
\multirow{8}{*}{\rotatebox[origin=c]{90}{Caen}} 
 & 0 & -0.6 & -1.7 & -13.1 & -19.4 & -1.7 & -19.4 & -1.7 \\ 
 & 1 & 13.4 & 6.4 & -6.1 & -17.3 & 6.4 & -17.3 & 6.4 \\ 
 & 2 & 29.0 & 5.6 & 4.9 & -12.3 & 5.6 & -12.3 & 5.6 \\ 
 & 3 & 35.6 & 2.3 & 13.2 & -6.7 & 2.3 & -6.7 & 2.3 \\ 
 & 4 & 37.7 & -0.4 & 19.0 & -1.5 & -0.4 & -1.5 & -0.4 \\ 
 & 5 & 39.5 & -0.5 & 24.7 & 5.0 & -0.5 & 5.0 & -0.5 \\ 
 & 6 & 26.2 & -8.8 & 22.9 & 11.7 & -8.8 & 11.7 & -8.8 \\ 
 & 7 & 28.7 & 2.5 & 28.1 & 23.1 & 2.5 & 23.1 & 2.5 \\ 
\\\multirow{3}{*}{\rotatebox[origin=c]{90}{Chid.}} 
 & 0 & 25.0 & 24.0 & 10.4 & -6.0 & 24.0 & -6.0 & 24.0 \\ 
 & 1 & 11.9 & 9.5 & -2.6 & -11.3 & 9.5 & -11.3 & 9.5 \\ 
 & 2 & -3.1 & -3.9 & -15.2 & -17.6 & -3.9 & -17.6 & -3.9 \\ 
\\\multirow{3}{*}{\rotatebox[origin=c]{90}{Ferg.}} 
 & 0 & 48.8 & 17.2 & 28.6 & -1.1 & 17.2 & -1.1 & 17.2 \\ 
 & 1 & 34.7 & 4.4 & 30.9 & 8.6 & 4.4 & 8.6 & 4.4 \\ 
 & 2 & 14.0 & 12.4 & 14.0 & 10.4 & 12.4 & 10.4 & 12.4 \\ 
\\\multirow{12}{*}{\rotatebox[origin=c]{90}{Weigend}} 
 & 0 &  &  &  &  &  & -19.5 & 3.3 \\ 
 & 1 &  &  &  &  &  & -2.7 & 4.0 \\ 
 & 2 &  &  &  &  &  & 2.6 & -0.4 \\ 
 & 3 &  &  &  &  &  & -20.6 & -2.5 \\ 
 & 4 &  &  &  &  &  & -6.3 & -3.5 \\ 
 & 5 &  &  &  &  &  & 5.3 & -3.8 \\ 
 & 6 &  &  &  &  &  & -6.5 & 4.8 \\ 
 & 7 &  &  &  &  &  & 6.3 & 2.0 \\ 
 & 8 &  &  &  &  &  & 13.6 & 0.9 \\ 
 & 9 &  &  &  &  &  & -9.6 & 0.5 \\ 
 & 10 &  &  &  &  &  & 11.2 & 6.1 \\ 
 & 11 &  &  &  &  &  & 13.3 & -2.6 \\ 
\midrule\addlinespace[2mm] 
\multicolumn{2}{c}{MAE}  & 24.87\tnote{*} & 7.11 & 17.23\tnote{*} & 15.06\tnote{*} & 7.07 &  & \\ 
\multicolumn{2}{c}{$\pm$ SD} & \pm 14.35 & \pm 6.83 & \pm 9.34 & \pm 12.19 & \pm 7.17 &  &  \\ \\ 
\multicolumn{2}{c}{RMSE} & 28.46 \tnote{*}& 9.69 & 19.48 \tnote{*}& 19.17 \tnote{*}& 9.94 &  &  \\ \\ 
\multicolumn{2}{c}{AIC} &  &  &  &  &  & 181.03 & 151.85 \\ 
\bottomrule \end{tabular} 
\begin{tablenotes} 
\item[$*$] significantly different to \gls{hyd-weig} predictions
\end{tablenotes} 
\end{threeparttable} 
\end{adjustwidth} 
\end{table}

\Cref{tab:error_summary} summarizes the prediction errors of the competing models and resulting metric scores on our investigated data sets. \gls*{RMSE} and \gls*{MAE} were defined as the metrics to assess predictive accuracy. Their \gls*{MAE} scores were 24.87 with a standard deviation of absolute errors (SD) of 14.35 for \gls*{w'bal-bart} and 7.11(SD=6.83) for \gls*{hyd-weig} ($\mathrm{p}<.001$ for the difference in \gls*{MAE}s, bootstrap hypothesis test). The \gls*{RMSE} scores on Caen, Chidnok, and Ferguson data sets were 28.46 for \gls*{w'bal-bart} and 9.69 for \gls*{hyd-weig}. Also the bootstrap hypothesis test with the absolute difference in \gls*{RMSE}s as its test statistic resulted in $\mathrm{p}<.001$. 

Both remaining models \gls*{w'bal-skib} and \gls*{w'bal-weig} could be compared to \gls*{hyd-weig} on the Bartram, Caen, Chidnok, and Ferguson data sets. The \gls*{hyd-weig} featured the lowest \gls*{MAE} with 7.07, the lowest SD with 7.17, and lowest \gls*{RMSE} with 9.94. \Gls*{w'bal-skib} predictions were significantly different to \gls*{hyd-weig} ($\mathrm{p}<.001$ with the \gls*{MAE} test statistic and $\mathrm{p}=.001$ with \gls*{RMSE}). \Gls*{w'bal-weig} predictions were significantly different to \gls*{hyd-weig} ($\mathrm{p}=.019$ with the \gls*{MAE} test statistic and $\mathrm{p}=.031$ with \gls*{RMSE}). 

\gls*{AIC} was chosen as the metric to assess which model provides the best trade-off between predictive capabilities and complexity. Models must be fitted to and tested on the same data for \gls*{AIC} scores to be comparable. Hence, as reflected in the last two columns of \Cref{tab:error_summary}, \gls*{w'bal-weig} and \gls*{hyd-weig} could be compared on combined data points of all covered data sets. With a $k$ of 3 for \gls*{w'bal-weig} and a $k$ of 8 for \gls*{hyd-weig} the resulting scores were 151.85 for \gls*{hyd-weig} and 181.03 for \gls*{w'bal-weig}. The \gls*{hyd-weig} achieved the lower \gls*{AIC} score.

\newpage
\section{Discussion}
\label{sec:discussion}

In this study, we compared the prediction capabilities and goodness of fit of \gls*{hyd-weig} to that of \gls*{w'bal} models. We hypothesized that the hydraulic model would more accuratly predict observed recovery ratios observed in past studies. Models were compared on extracted data from five studies and the \gls*{hyd-weig} model outperformed the \gls*{w'bal-skib}, \gls*{w'bal-bart}, and \gls*{w'bal-weig} models with respect to objective \gls*{RMSE}, \gls*{MAE}, and \gls*{AIC} metrics. Our findings therefore support the hypothesis. 
We discuss below our results in more detail and interpret them in context of findings of previous literature. We present arguments for why the \gls*{hyd-weig} outperformed the \gls*{w'bal-ode} models and we propose limitations and future work. Finally, we end this section with statements about significance and implications of our results.

\subsection{Interpretation and contextualization}

We observed that the standard deviations of absolute prediction errors in \Cref{subsec:metrics_summary} as well as the overall \gls*{MAE} and \gls*{RMSE} were considerably lower for the \gls*{hyd-weig} than for the \gls*{w'bal-ode} models. But when averaging the prediction errors on isolated data sets listed in \Cref{tab:error_summary}, \gls*{hyd-weig} only made more accurate predictions than its competitors on the Bartram and Caen data sets. For the Bartram data set, the \gls*{MAE} of \gls*{hyd-weig} was 6.96, compared to 18.74 for \gls*{w'bal-skib}, and 26.82 for \gls*{w'bal-weig} respectively. For the Caen data set, the \gls*{MAE} of \gls*{hyd-weig} was the lowest with 3.52. On the remaining Chidnok data set it was \gls*{w'bal-skib} that achieved the lowest \gls*{MAE} with 9.4 and on the Ferguson data set it was \gls*{w'bal-weig} with 6.7.

As pointed out by \citet{skiba_w_2021} and \cite{sreedhara_survey_2019}, \gls*{w'bal-ode} models are meant to be applied to any athlete on a wide range of possible conditions. A lower \gls*{MAE} score for $W^\prime_\mathrm{weig}$ on the Ferguson data set means $W^\prime_\mathrm{weig}$ predicted recovery ratios more closely for the particular group (six recreational active men) under the particular test conditions that Ferguson tested. However, to determine the usefulness of a model for predicting performance for high-intensity intermittent exercise in a more general sense, models have to be evaluated on a multitude of scenarios. After combining all data sets, $\mathrm{hydraulic}_\mathrm{weig}$ achieved the overall lowest \gls*{MAE} score, which means that $\mathrm{hydraulic}_\mathrm{weig}$ could predict recovery ratios overall more accurately for a range of groups and settings. 

\begin{figure}
    \centering
    \includegraphics[width=\linewidth]{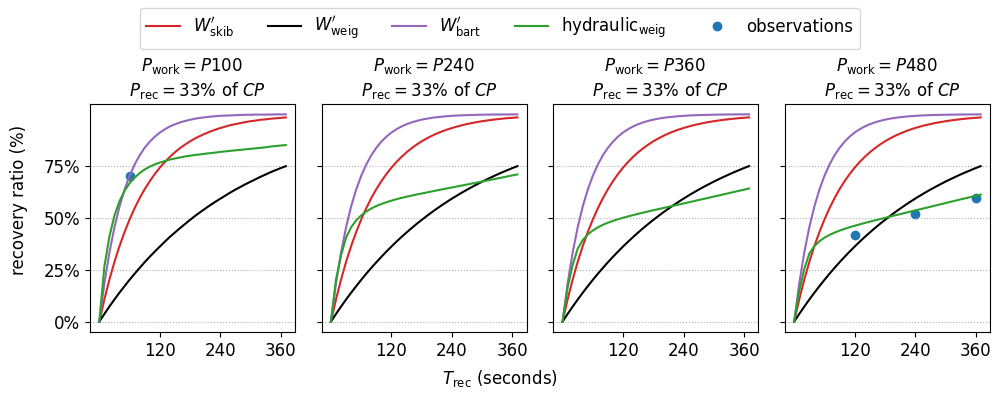}
    \caption{Simulated recovery ratios using the \gls*{hyd-weig} and \gls*{w'bal-ode} models in response to prior exercise of differing intensities. The plots show that \gls*{w'bal-ode} models are insensitive to the properties of prior exhausting exercise, i.e., their predictions were not affected by $P_{\mathrm{work}}$. In contrast, the \gls*{hyd-weig} was sensitive to the prior exercise properties. Performance models \gls*{w'bal-ode} were configured with a \gls*{cp} = 393, \gls{w'} = 23300 and \gls*{hyd-weig} featured the configuration $[$23111.91, 65845.28, 391.57, 148.88, 24.15,  0.73, 0.01, 0.24$]$. All simulations differed only in $P_{\mathrm{work}}$, which decreased from $P100$ to $P480$ in the simulations depicted from left to right. $P_{\mathrm{work}}$ = $P100$, prescribed by \citet{bartram_accuracy_2018}, was the highest intensity out of compared studies. They investigated recovery after 60 seconds, therefore the \gls*{w'bal-bart} model prediction after 60 seconds is marked as the observation. $P_{\mathrm{work}}$ = $P480$ was the lowest prescribed intensity out of compared studies and recovery ratios for $P_{\mathrm{work}}$ = $P480$, $P_{\mathrm{rec}}$ = $33\%$ of \gls*{cp} from the Weigend data set were marked as observations on the right.} \label{fig:pwork_mult}
\end{figure}

The less consistent prediction quality across data sets of the \gls*{w'bal-ode} models agrees with findings by \citet{caen_reconstitution_2019}, who proposed that the predictive capabilities of \gls{w'bal} models may improve with modifications that account for intensity and duration of prior exhaustive exercise. As an example, out of all compared studies in this work, \cite{bartram_accuracy_2018} prescribed the highest work bout intensity for their experimental setup ($P_\mathrm{work}$ = $P100$). Considering the suggestion by \citet{caen_reconstitution_2019} that a shorter time to exhaustion at a high intensity allows a quicker recovery, it seems reasonable that the \gls*{w'bal-bart} model estimated the fastest recovery kinetics out of all recovery models. 

Conversely, the \citet{caen_reconstitution_2019} study prescribed the lowest work bout intensity out of all compared studies ($P_\mathrm{work}$ = $P480$). Their observed recovery ratios are summarized in the Weigend data set and were slower than the \gls*{w'bal-bart} predictions. This observation again matches the assumption that a longer exhaustive exercise at a lower intensity requires a longer recovery. 

Despite the differences in observed recovery rates, the \gls*{w'bal-ode} models allow for only a single recovery rate no matter the nature of the prior exercise. To illustrate this point, we conducted simulations to depict the influence of prior exercise intensity on the recovery ratios predicted by the \gls*{w'bal-ode} and \gls*{hyd-weig}. \Cref{fig:pwork_mult} depicts four simulations. All simulations shared the same test setup except for differing $P_\mathrm{work}$ intensities. The simulation on the left had $P_\mathrm{work}$ = $P100$ as prescribed by \cite{bartram_accuracy_2018}, the simulation on the right featured the lowest $P_\mathrm{work}$ = $P480$ as found in the Weigend data set. From left to right, $P_\mathrm{work}$ of the simulations decreased step wise. \cite{bartram_accuracy_2018} investigated recovery after 60 seconds and therefore the \gls*{w'bal-bart} prediction after 60 seconds is marked as the observation on the left. Recovery ratios with $P_\mathrm{work}$ = $P480$ and $P_\mathrm{rec}$ = $33\%$ of $CP$ of the Weigend data set are marked as observations on the right. The recovery ratios predicted by the \gls*{w'bal-ode} models were the same for each $P_{\mathrm{work}}$ and their predictions were therefore unable to fit all observations equally well. In contrast, the hydraulic model could account for such characteristics. 

This result occurred because of the interactions between the three tanks that \gls*{hyd-weig} uses to model energy recovery. For example, during high-intensity exercise, the liquid level in $AnF$ would rapidly decrease and the contribution of $AnS$ would be less than during lower-intensity exercise when the liquid level in $AnF$ would decrease more slowly. Differences in fill states of $AnS$ affected recovery estimations and enabled \gls*{hyd-weig} to predict rapid recovery after high-intensity exercise and a slower recovery after exercise at a lower intensity. We suggest that standard deviations of \gls*{MAE} as well as overall \gls*{MAE} and \gls*{RMSE} scores of \gls*{hyd-weig} model were smaller than those of \gls*{w'bal-ode} models because the hydraulic model could account for characteristics of prior exhaustive exercise.

Further, we propose that \gls*{hyd-weig} has also achieved overall better metric scores because it better captured the bi-exponential nature of energy recovery, as opposed to the mono-exponential \gls*{w'bal-ode} models. Indeed, the observed recovery ratios of the Caen data set increased rapidly from 0 seconds to 120 seconds and then continued to rise more slowly at longer durations (\Cref{fig:caen_2021_rec}). \cite{caen_w_2021} showed that their observations were well explained with a bi-exponential model that implements a steeper slope during the beginning of recovery. Also, the first \gls*{w'bal} paper by \citet{skiba_modeling_2012} proposed an alternative bi-exponential version of their \gls*{w'bal} model with two $\Tau_t$s but such bi-exponential \gls*{w'bal} models have yet to be applied in practice. 

\subsection{Limitations and future work}
\label{subsec:limitations}

The observed improved prediction quality of \gls*{hyd-weig} comes at the cost of a time-demanding fitting process. As outlined in \Cref{subsec:hyd_model}, our fitting process from \citet{weigend_new_2021} requires \gls*{cp} and \gls*{w'} of an athlete as inputs and then obtains fitted \gls*{hyd-weig} configurations using an evolutionary computation approach. Different \gls*{cp} and \gls*{w'} values require a new \gls*{hyd-weig} configuration to be fitted to them. Additionally, fittings for our comparison took computation times of 3 hours or more on 7 cores of an $\text{Intel}^\text{\textregistered}$ $\text{Xeon}^\text{\textregistered}$ CPU E5-2650 v4 @ 2.20GHz each. On the other hand, obtaining $\Tau_t$ for our \gls*{w'bal-ode} models was solved in milliseconds and they can be applied to any \gls*{cp} and \gls*{w'} combination without fitting $\Tau_t$ anew. Therefore, the application of \gls*{hyd-weig} is a time consuming task in comparison to the application of \gls*{w'bal-ode} models. In order to improve the feasibility of application, future work must optimize the hydraulic model fitting process to minimize this limitation.

Further, prediction results show that estimated recovery ratios of the used \mbox{WB1 $\rightarrow$ RB $\rightarrow$ WB2} protocol are affected by rounding errors that arose from the recovery ratios being estimated from simulations in discrete time steps. For example, in \Cref{tab:weigend_comp}, it was observed that the \gls*{w'bal-ode} model predictions between the $P240$ and $P480$ trials varied slightly in a few cases, even though these the recovery kinetics should have been the same. The variations were caused by simulation time steps of size 0.1 that made rounding effects play a bigger role in shorter work bouts of $P240$. Smaller step sizes would decrease the error, but to more fully prevent inaccuracies, future work ought to formalize model simulations in differential equations that don't require estimations in discrete time steps.

Predicted recovery ratios of the \gls{hyd-weig} at recovery intensities close to \gls*{cp} require further investigation too. The comparison on the $D_{CP}$0 case of the Bartram data set (See \Cref{fig:bartram_study}) revealed that \gls*{hyd-weig} predicts a slight recovery during exercise at \gls*{cp}. No liquid from $Ae$ flowed back into the system but the ongoing flow from $AnS$ to $AnF$ still caused liquid level in $AnF$ to rise during the recovery bout. Recovery while exercising at \gls*{cp} intensity is a controversial assumption that is made to an even stronger extent by the original \gls*{w'bal} model of \citet{skiba_modeling_2012}. Such dynamics have to be taken into consideration when the models are used for predictions and are important directions for future investigation.

Additionally, extracted recovery ratio observations from previous studies come with associated uncertainties that could not be considered in our \gls*{MAE}, \gls*{RMSE} and \gls*{AIC} scores. As an example, the standard deviations of observed recovery ratios by \citet{caen_w_2021} depicted in \Cref{fig:caen_2021_rec} are greater than reported standard deviations by \citet{ferguson_effect_2010} depicted in \Cref{fig:ferguson_study}. One could argue that comparisons to reported means by \citet{ferguson_effect_2010} therefore provide a better indication of prediction quality. Unfortunately, we could not incorporate these standard deviations into goodness-of-fit metrics because of how different recovery ratios were reported in compared studies. \citet{caen_w_2021} and \citet{ferguson_effect_2010} reported averaged observed recovery ratios with standard deviations, for the Bartram data set we had to use \gls*{w'bal-bart} predictions as observations to compare to, in \citet{weigend_new_2021} we derived our values from \citet{caen_reconstitution_2019} without standard deviations, and for the Chidnok data set we fitted constant values for $\Tau_t$ for \gls*{w'bal-ode} models to their reported times to exhaustion to obtain comparable recovery ratios in percent. We believe these various formats of observations highlight the need for more and more comparable studies on energy recovery dynamics.

Larger data sets are vital for more educated investigations of recovery models and their improvement in future work. We see the combination of data sets in this work as a step towards this direction. In order to improve and compare models more holistically, it is important that more comparable studies are conducted in the future and combined into a larger test bed for performance models.

\subsection{Significance and implications}

To the best of our knowledge, performance models on energy recovery during intermittent exercise have yet to be compared in such detail. Our comparison on data from five studies allowed a more holistic view on recovery dynamics and confirmed limitations of \gls*{w'bal-ode} models that were suggested by previous literature. Our results imply that more complex models like \gls*{hyd-weig} can improve energy recovery predictions. We propose that further efforts to merge and compare data are significant steps to bring the research area of energy recovery modeling forward. We further propose that the predictive capabilities of hydraulic models look strong and that hydraulic models have to be considered as a possible future direction to advance energy recovery modeling.

\section{Conclusion}
\label{sec:conclusion}

We conclude that the \gls*{hyd-weig} outperformed \gls*{w'bal-ode} models when fit to multiple independent data sets featuring intermittent high-intensity exercise. The predictive accuracy and goodness of fit was better for the hydraulic model, even for the \gls*{AIC} metric, which includes a penalty for the number of model parameters. The hydraulic model is thus likely more generalizable than \gls*{w'bal-ode} models, which are typically applied within narrow contexts. Future research should focus on improving the feasibility of the hydraulic model, because it is computationally more burdensome to use than \gls*{w'bal-ode} models. Pending such improvements, we foresee athletes adopting this model to optimize pacing and interval training workouts. To contribute towards further advancements we publish all material, extracted data, and simulation scripts here: \url{https://github.com/faweigend/pypermod}.

\section*{Declarations}
\paragraph{Funding}
No funding was received to assist with the preparation of this manuscript.
\paragraph{Conflict of interest}
The authors have no conflicts of interest to declare that are relevant to the content of this article.
\paragraph{Availability of data and material}
All material and extracted data are summarized in \Cref{tab:bart_comp,tab:caen_comp,tab:chid_comp,tab:ferg_comp,tab:weigend_comp} and are also available here: \url{https://github.com/faweigend/pypermod}.
\paragraph{Code availability}
All code is thoroughly documented and available as source code and as a python package here: \url{https://github.com/faweigend/pypermod}.

\bibliographystyle{apalike}      
\bibliography{references}   

\end{document}

%% file: acronyms.tex
\newacronym{cp}{\ensuremath{CP}}{critical power}
\newacronym{p}{\ensuremath{P}}{constant power output}
\newacronym{w'}{\ensuremath{W^\prime}}{finite energy reserve for work above critical power}
\newacronym{w'bal}{\ensuremath{W^\prime_\mathrm{bal}}}{\ensuremath{W^\prime} balance}
\newacronym{w'bal-ode}{\ensuremath{W^\prime_\mathrm{bal-ode}}}{\gls*{w'bal} Ordinary Differential Equation}

\newacronym{tau-skib}{\ensuremath{\Tau_{\mathrm{skib}_t}}}{\ensuremath{\Tau_{t}} of the \gls*{w'bal-ode} model by \cite{skiba_intramuscular_2015}}
\newacronym{tau-bart}{\ensuremath{\Tau_{\mathrm{bart}_t}}}{\ensuremath{\Tau_{t}} by \citet{bartram_accuracy_2018}}
\newacronym{tau-weig}{\ensuremath{\Tau_{\mathrm{weig}_t}}}{\ensuremath{\Tau_{t}} created on the same recovery measures that were used to fit \gls*{hyd-weig}}
\newacronym{hyd-weig}{\ensuremath{\mathrm{hydraulic}_\mathrm{weig}}}{hydraulic M-M model by \citet{morton_three_1986} as interpreted by \citet{weigend_new_2021}}

\newacronym{w'bal-skib}{\ensuremath{W^\prime_\mathrm{skib}}}{\gls*{w'bal-ode} model with time constant fitted by \cite{skiba_intramuscular_2015}}
\newacronym{w'bal-bart}{\ensuremath{W^\prime_\mathrm{bart}}}{\gls*{w'bal-ode} model with time constant fitted by \cite{bartram_accuracy_2018}}
\newacronym{w'bal-weig}{\ensuremath{W^\prime_\mathrm{weig}}}{\gls*{w'bal-ode} model with \gls*{tau-weig}}

\newacronym{tte}{\text{TTE}}{time to exhaustion}
\newacronym{RMSE}{RMSE}{Root Mean Squared Error}
\newacronym{MAE}{MAE}{Mean Absolute Error}
\newacronym{AIC}{\ensuremath{\mathrm{AIC}_\mathrm{c}}}{the small-sample version of the Akaike Information Criterion}